\documentclass[twoside]{article}
\usepackage{qic-lucio,epsfig}
\usepackage{amssymb}

\renewcommand{\thefootnote}{\fnsymbol{footnote}}  

\newtheorem{remark}{Remark}

\begin{document}
\setlength{\textheight}{8.0truein}    

\runninghead{Controlling the coherence  $\ldots$}
            {L\'{u}cio Fassarella}

\normalsize\textlineskip
\thispagestyle{empty}
\setcounter{page}{1}


\vspace*{0.88truein}

\alphfootnote

\fpage{1}

\centerline{\bf CONTROLLING THE COHERENCE IN A PURE DEPHASING MODEL}
\vspace*{0.035truein}
\centerline{\bf FOR AN ARBITRARILY PRESCRIBED TIME SPAN}
\vspace*{0.37truein}
\centerline{\footnotesize
L\'{U}CIO FASSARELLA
}
\vspace*{0.015truein}
\centerline{\footnotesize\it Departamento de Matem\'{a}tica Aplicada, Universidade Federal do Esp\'{i}rito Santo, Rodovia BR 101 Norte, Km60.}
\baselineskip=10pt
\centerline{\footnotesize\it{S\~{a}o Mateus - ES.  29932-540, Brazil}\footnote{E-mail: lucio.fassarella@ufes.br.}}
\vspace*{0.21truein}
\abstracts{
We present an open-loop unitary strategy to control the coherence in a pure dephasing model (related to the phase-flip channel) that is able to recover, for whatever prescribed time span, the initial coherence at the end of the control process. The strategy's key idea is to steer the quantum state to the subset of invariant states and keep it there the necessary time, using a fine tuned control Hamiltonian.}{}{}
\vspace*{10pt}
\vspace*{3pt}

\vspace*{1pt}\textlineskip    
\section{Introduction}        
The open-loop unitary controlling is an important methodology of quantum control, having the characteristic of avoiding totally any perturbation of systems during the control process, feature that simplifies the technological apparatus required to implement the control in practice. In spite of its limitations, it has a vast range of applications, including
quantum chemistry, quantum optics, quantum information and also biophysics.

The unitary control of Markovian quantum systems is strongly motivated
because the Markovian approximation can be used to describe a wide class of
open quantum systems (besides the closed ones), enabling the theory to be
used in many practical problems \cite{BP2003, AL2007}. Such systems are also particularly amenable because their dynamics can be suitably transformed into real linear dynamical systems, through coherent vector representation \cite[pp.50-57]{AL2007}.

Finally, the control of coherence in quantum systems is a demanding task for
the development of quantum information and computation technologies, fact
evidenced by the vast literature on the subject -- see \cite{BH2002, Alt2004, Al2006, BG2013} and references quoted therein. This subject has been massively studied but there are many open questions even
in the most simple situations. For example, the unitary tracking-control strategy to stabilize (keep
constant) the coherence  of a pure dephasing model presented in \cite{LS2005} suffers from a severe limitation,
unavoidable for all unitary control strategies which stabilize the coherence in this model
(whether performed in a closed-loop or in an open-loop fashion): the control can
be carried out only within a finite time span, at the end of which the control
fields diverge. Nevertheless, \emph{it's possible to control the quantum state in order to recover the initial coherence after an arbitrary prescribed time span if one is allowed to use control Hamiltonians that don't keep the coherence constant (necessarily)}. The contribution of this paper is twofold: the definition of a general strategy to find a fine tuned control Hamiltonian to recover the coherence of a given initial state after any prescribed time span, and the explicit application of such strategy in the model just mentioned, called here \textit{dephasing qubit}.

The structure of the paper is simple. In Section \ref{section_definitions} we review basic concepts in order to give a short and precise formulation of our problem in Section \ref{section_problem}. In Section \ref{section_solution} we define a general strategy to tackle such kind of problem, we apply it to solve the specific problem stated previously and give a numerical example. In the final Section \ref%
{section_conclusion} we discuss our results and comment related issues. The Appendix A focuses the concept of \emph{limit time}, related to the definition of the control Hamiltonian.
\section{Dephasing qubit}
\label{section_definitions}
\noindent
\setcounter{footnote}{0}
\renewcommand{\thefootnote}{\alph{footnote}}

We start recalling basic definitions and results concerning the \textit{dephasing qubit} model, using a notation borrowed from \cite{LS2005}.

A general quantum state (density matrix) of a qubit can be written in terms of the \emph{identity operator} $I$ and \emph{%
Pauli matrices} ($\sigma _{x}, \sigma _{y}, \sigma _{z}$), whose coefficients define the so called \emph{Bloch vector}:%
\begin{equation}
\rho =\frac{1}{2}\left( I+v_{x}\sigma _{x}+v_{y}\sigma _{y}+v_{z}\sigma
_{z}\right) ,\ v=\left( v_{x},v_{y},v_{z}\right) \in B:=\left\{ v\in \mathbb{%
R}^{3};\ \left\Vert v\right\Vert \leq 1\right\}.
\label{model-pd_density-matrix}
\end{equation}%
The purity and coherence are defined, respectively, by%
\begin{equation}
P\left( \rho \right) :=v_{x}^{2}+v_{y}^{2}+v_{z}^{2}\ ,\ C\left( \rho
\right) :=v_{x}^{2}+v_{y}^{2}.  \label{model-pd_coherence}
\end{equation}%
The \emph{free dynamics} is given by the master equation 
\begin{equation}
\frac{d}{d t}\rho \left( t \right)=\frac{\gamma }{2}\left( \sigma _{z}\rho \left( t \right)
\sigma _{z}-\rho \left( t \right) \right) ,  \label{model-pd_dynamics}
\end{equation}%
where $\gamma >0$ is a damping coefficient. A control Hamiltonian 
\begin{equation}
H\left( t\right) =\frac{1}{2}\left( u_{1} \left( t \right) \sigma _{x}+u_{2} \left( t \right) \sigma
_{y}+u_{3} \left( t \right) \sigma _{z}\right) ;\ u_{1}\left( t\right) ,u_{2}\left( t\right)
,u_{3}\left( t\right) \in \mathbb{R}^{3} .
\label{model-pd_control-hamiltonian}
\end{equation}%
affects the free dynamics according with%
\begin{equation}
\frac{d}{dt}\rho \left( t\right) =\frac{\gamma }{2}\left( \sigma _{z}\rho
\left( t\right) \sigma _{z}-\rho \left( t\right) \right) -i\left[ H\left(
t\right) ,\rho \left( t\right) \right] .
\label{model-pd_dynamics-controlled}
\end{equation}%

The model has a \textit{decoherence-free subset of states}\footnote{We use the term \textquotedblleft\textit{decoherence-free subset of states}\textquotedblright \ to distinguish it from the related concept of \textit{decoherence free subspaces}, for which we refer to \cite{LW2003, Li2014}.} \ defined by:
\[
V_{z}:=\left\{ \frac{1}{2} \left( I+\xi \sigma _{z} \right);\ -1\leq \xi \leq 1\right\}. 
\label{model-pd_dfs}
\]
For time-dependent states evolving within $V_{z}$, the dynamics is reduced to the Liouville-von Neumann equation (meaning that its time evolution is unitary):
\[
\frac{d}{dt}\rho \left( t\right) = -i\left[ H\left(
t\right) ,\rho \left( t\right) \right],  \ \mbox{if} \  \rho \left( t\right) \in V_{z}. 
\label{model-pd_ln}
\]
Equation (\ref{model-pd_dynamics-controlled}) turns out to be equivalent to the following system for the Bloch
vector's coordinates:%
\begin{equation}
\left\{ 
\begin{array}{l}
2\dot{v}_{x}=-\gamma v_{x}+u_{y}v_{z}-u_{z}v_{y} \\ 
2\dot{v}_{y}=-\gamma v_{y}-u_{x}v_{z}+u_{z}v_{x} \\ 
2\dot{v}_{z}=u_{x}v_{y}-u_{y}v_{x}.%
\end{array}%
\right.  \label{ucontrol_dynamics}
\end{equation}

Given the values of purity and coherence of an initial state $\rho \left(
0\right) $, 
\begin{equation}
p=v_{x}^{2}\left( 0\right) +v_{y}^{2}\left( 0\right) +v_{z}^{2}\left(
0\right) ,\ \ c=v_{x}^{2}\left( 0\right) +v_{y}^{2}\left( 0\right) ,
\label{model_pd_pc0}
\end{equation}%
the corresponding \emph{breakdown time} is defined by%
\begin{equation}
t_{b}:=\frac{p-c}{\gamma c} .  \label{model_pd_breakdown-time}
\end{equation}

\vspace*{12pt}
\noindent
\begin{theorem} In the \textit{dephasing qubit}, the coherence of a time-dependent state cannot be stabilized (kept constant) by unitary controlling for a time span greater then the \textit{breakdown time} Eq. (\ref{model_pd_breakdown-time}).\\
\begin{proof} \noindent Let $H\left(t \right)$ be the Hamiltonian of a unitary control and let $\rho \left( t\right)$ be a solution of the system (\ref%
{model-pd_dynamics-controlled})\ having constant coherence, $C\left( \rho
\left( t\right) \right) =C\left( \rho \left( 0\right) \right) =c$. Assume
that $H \left( t\right) $ and $\rho \left( t\right) $ are defined for $t\in \left[ 0,T\right] $, for
some $T>0$. The dynamical equations for the Bloch coordinates (\ref%
{ucontrol_dynamics}) imply%
\begin{equation}
\frac{d}{dt}\left( v_{x}^{2}+v_{y}^{2}+v_{z}^{2}\right) =-\gamma \left(
v_{x}^{2}+v_{y}^{2}\right) .  \label{model-pd_boch-vector_purity-decay}
\end{equation}%
So, the coherence (given by Eq. (\ref{model-pd_coherence})) is kept invariant if, and only if,%
\begin{equation}
\frac{d}{dt}\left( v_{x}^{2}+v_{y}^{2}\right) =0, \ \frac{d}{dt}v_{z}^{2}=-\gamma \left( v_{x}^{2}+v_{y}^{2}\right) .   \label{model_pd_cv1}
\end{equation}%
In this case, it follows that
\[
v_{z}^{2}\left( t\right) =v_{z}^{2}\left( 0\right) -c\gamma t,\ \forall t\in 
\left[ 0,T\right].
\label{model_pd_cv2}
\]%
Since $v_{z}\left( t\right) $ must be real and $v_{z}^{2}\left( 0\right) =p-c
$, the condition $v_{z}^{2}\left( t\right) \geq 0$ implies that $T\leq t_{b}$; this means that the time span
$H \left( t\right) $ and $\rho \left( t\right) $ are defined cannot be greater than the breakdown
time.$\Box$
\end{proof}
\end{theorem}

Due to \emph{Theorem 1}, to recover the coherence of an initial state after a time span greater than the \textit{breakdown time} one must accomplish control strategies that do not keep constant the coherence; so, it is worthwhile to consider the problem formulated in the next section.

\section{The Problem}
\label{section_problem}
\noindent

\begin{quotation}
\textbf{\ \textit{Problem}:} In the \textit{dephasing qubit}, for a given $T>0$ and initial state $\rho \left(
0\right) $, set a control Hamiltonian to steer the state's evolution according
with Eq. (\ref{model-pd_dynamics-controlled}) in such a way that the coherence of the system' state after the time span $T$ turns out to be
equal to the coherence of the initial state, \textit{i.e.},%
\[
C\left( \rho \left( T\right) \right) =C\left( \rho \left( 0\right) \right) .
\]
\end{quotation}

\begin{remark} \label{remark1}
This problem cannot be solved using only unitary controlling if coherence and purity start equal: according with Theorem 1 and Eq. (\ref{model_pd_breakdown-time}), nothing can be done in this way if $v_{z}\left(0\right) =0$. Also, the same theorem and equation imply that there is nothing to do if the initial coherence is zero.
\end{remark}

\section{The Solution\label{section_solution}}
\noindent

In this section, we define and apply a simple and general strategy to solve the specific \emph{Problem} previously stated. This strategy uses the \emph{decoherence-free subset of states} of the \textit{dephasing qubit}.

To simplify the calculations, we deal first with a special initial state and then generalize the result. After due developments, the solution will be presented in the form of an algorithm.

\subsection{The Strategy \label{section_strategy}}
\noindent

\begin{quotation}  

(i) first, steer\ the qubit's state to the \emph{decoherence-free subset of states};

(ii) second, keep the state within $V_{z}$ for the period needed;

(iii) finally, bring the system to some final state which has coherence equals to the initial value at the end of the process.
\end{quotation}

\subsection{Solving the Problem for special initial state}
\label{section_special-case}
\noindent

Consider an initial state $\rho \left( 0\right)$ with purity $p$ greater than a positive coherence $c$ which has the following special form\footnote{
As we already have said: if $v_{x}\left( 0\right) =0$, there is nothing to be done; if $v_{z}\left(
0\right) =0$, there is nothing which can be done.}
\begin{equation}
\rho \left( 0\right) =\frac{1}{2}I+\frac{1}{2}v_{x}\left( 0\right) \sigma
_{x}+\frac{1}{2}v_{z}\left( 0\right) \sigma _{z},\ v_{z}\left( 0\right) \neq
0 < v_{x}\left( 0\right) .  \label{ucontrol_state-r0}
\end{equation}
In this case, we can use control fields having $y$-component being the only
nonzero -- a choice that confines the time-dependent Bloch vector to the $xz$%
-plane during its entire evolution:%
\begin{equation}
u_{x}=0=u_{z};\ u_{y}=:\epsilon u,\ \epsilon =\pm 1,\ u>0.
\label{ucontrol_control-field}
\end{equation}%
For convenience we have introduced the signal $\epsilon $ which determines
de direction the state's Bloch vector rotates in the $xz$-plane due to the
action of the control Hamiltonian: $\epsilon =+1$\ corresponds to clockwise
direction and $\epsilon =-1$\ corresponds to counterclockwise direction.

\bigskip

The dynamics of Bloch vector Eq. (\ref{ucontrol_dynamics}) under action of the
control fields Eq. (\ref{ucontrol_control-field}) with initial conditions at $%
t_{0}$ added turns to%
\begin{equation}
\left\{ 
\begin{array}{l}
2\dot{v}_{x}=-\gamma v_{x}+\epsilon uv_{z} \\ 
2\dot{v}_{y}=-\gamma v_{y} \\ 
2\dot{v}_{z}=-\epsilon uv_{x} \\ 
v_{y}\left( t_{0}\right) =0,\ v_{x}^{2}\left( t_{0}\right) +v_{z}^{2}\left(
t_{0}\right) \leq 1.%
\end{array}%
\right.  \label{ucontrol_dynamics-3}
\end{equation}%
Assuming the control field $u$ to be constant and%
\begin{equation}
u>\frac{\gamma }{2},  \label{ucontrol_uasumption}
\end{equation}%
the solution of Eq. (\ref{ucontrol_dynamics-3}) is given by%
\begin{equation}
\left\{ 
\begin{array}{l}
v_{x}\left( t\right) =e^{-\gamma \left( t-t_{0}\right) /4}\left( v_{x}\left(
t_{0}\right) \cos \left( \frac{1}{4}\sqrt{4u^{2}-\gamma ^{2}}\left(
t-t_{0}\right) \right) +\right. \\ 
\ \ \ \ \ \ \ \ \ \ \ \ \ \ \ \ \ \ \ \ \ \left. +\frac{2\epsilon
uv_{z}\left( t_{0}\right) -\gamma v_{x}\left( t_{0}\right) }{\sqrt{%
4u^{2}-\gamma ^{2}}}\sin \left( \frac{1}{4}\sqrt{4u^{2}-\gamma ^{2}}\left(
t-t_{0}\right) \right) \right) \\ 
\  \\ 
v_{y}\left( t\right) =0 \\ 
\  \\ 
v_{z}\left( t\right) =e^{-\gamma \left( t-t_{0}\right) /4}\left( v_{z}\left(
t_{0}\right) \cos \left( \frac{1}{4}\sqrt{4u^{2}-\gamma ^{2}}\left(
t-t_{0}\right) \right) +\right. \\ 
\ \ \ \ \ \ \ \ \ \ \ \ \ \ \ \ \ \ \ \ \ \left. -\frac{2\epsilon
uv_{x}\left( t_{0}\right) -\gamma v_{z}\left( t_{0}\right) }{\sqrt{%
4u^{2}-\gamma ^{2}}}\sin \left( \frac{1}{4}\sqrt{4u^{2}-\gamma ^{2}}\left(
t-t_{0}\right) \right) \right) .%
\end{array}%
\right.  \label{ucontrol_dynamics-3_solution}
\end{equation}

Now, we describe separately the evolution of the controlled state $\rho
\left( t\right) $ during the first and the third stages of our control
process, starting from the initial state Eq. (\ref{ucontrol_state-r0}).

The shortest time span $\Delta t_{1}>0$ we need to steer $\rho \left( 0\right) $ to $V_{z}$ is given by the first positive zero of $v_{x}\left( t\right) $ in Eq. (\ref{ucontrol_dynamics-3_solution}) with $%
t_{0}=0$ and $t=\Delta t_{1}$; after some algebraic manipulation, we get $%
\Delta t_{1}$ explicitly:
\begin{equation}
\Delta t_{1}=\frac{4}{\sqrt{4u^{2}-\gamma ^{2}}}\arctan \left( \frac{\sqrt{%
4u^{2}-\gamma ^{2}}v_{x}\left( 0\right) }{\gamma v_{x}\left( 0\right)
-2\epsilon _{1}uv_{z}\left( 0\right) }\right) .
\label{ucontrol_dynamics-3_t1}
\end{equation}%
For $\Delta t_{1}$ to be positive, the argument of $\arctan $ in (\ref%
{ucontrol_dynamics-3_t1}) has to be positive, so we must set%
\begin{equation}
\epsilon _{1}:=-signal\left( v_{z}\left( 0\right) \right) . 
\label{ucontrol_usignal1}
\end{equation}%
Analogously, the shortest time span $\Delta t_{3}>0$ we need to steer $\rho \left( \Delta
t_{1}\right) $ from $V_{z}$ to some state having coherence equals to that of $\rho \left(
0\right) $, with the innocuous option to get the final state having its $\sigma _{x}$%
-component equals to that of $\rho \left( 0\right) $, is given by the first
positive solution of the following transcendent equation for $\Delta t_{3}$,
obtained from Eq. (\ref{ucontrol_dynamics-3_solution}) by setting $t_{0}=\Delta
t_{1}$, $t=\Delta t_{3}+\Delta t_{1}$ and $\epsilon _{3}=-\epsilon _{1}$:%
\begin{equation}
v_{x}\left( 0\right) =e^{-\gamma \Delta t_{3}/4}\left( \frac{2\epsilon
_{3}uv_{z}\left( \Delta t_{1}\right) }{\sqrt{4u^{2}-\gamma ^{2}}}\sin \left( 
\frac{1}{4}\sqrt{4u^{2}-\gamma ^{2}}\Delta t_{3}\right) \right) ,
\label{ucontrol_dynamics-3_t3}
\end{equation}%
where%
\[
v_{z}\left( \Delta t_{1}\right) = v_{z}\left( 0\right) e^{-\gamma \Delta t_{1}/4}\cos \left( \frac{1}{4}\sqrt{%
4u^{2}-\gamma ^{2}}\Delta t_{1}\right) \left( 1+\frac{%
\gamma v_{z}\left( 0\right) -2\epsilon uv_{x}\left( 0\right) }{\sqrt{%
4u^{2}-\gamma ^{2}}v_{z}\left( 0\right) }\tan \left( \frac{1}{4}\sqrt{%
4u^{2}-\gamma ^{2}}\Delta t_{1}\right) \right) . 
\]

\begin{remark}
Note the consistence of taking $\epsilon _{3}=signal\left( v_{x}\left(
0\right) v_{z}\left( 0\right) \right) $ in order for $\Delta t_{3}$ to be
positive in Eq. (\ref{ucontrol_dynamics-3_t3}), since $v_{z}\left( \Delta
t_{1}\right) $ has the same signal that $v_{z}\left( 0\right) $; this choice can be
verified by taking into account the definition of $\epsilon _{1}$ in Eq. (\ref%
{ucontrol_usignal1}).
\end{remark}

\bigskip

Now, to write down our control Hamiltonian which solves the \emph{Problem,}
we have to find a control field's intensity $u$ that guarantees the implicit
equation Eq. (\ref{ucontrol_dynamics-3_t3}) has a positive solution and such that%
\begin{equation}
T\geq \Delta t_{1}+\Delta t_{3}.  \label{model-pd_T}
\end{equation}%
This amounts to solve for $\Delta t_{1}$, $\Delta t_{3}$ and $u$ the system constituted by Equations (\ref
{ucontrol_dynamics-3_t1}) and (\ref{ucontrol_dynamics-3_t3}) and Inequality (%
\ref{model-pd_T}). Finally, by setting
\[
\epsilon = - signal\left( v_{z}\left( 0\right) \right), \ \Delta t_{2}:= T - \Delta t_{1}-\Delta t_{3}, 
\]
we can define the control Hamiltonian:
\begin{equation}
H\left( t\right) =\epsilon u\left[ \mathfrak{h}(\Delta
t_{1}-t)-\mathfrak{h}(t-\Delta t_{2})\right] \sigma _{y},\ 0\leq
t\leq T,
\label{control-hamiltonian-special}
\end{equation}
where $\mathfrak{h}$ denotes the Heaviside Step Function,
\[
\mathfrak{h}\left( t\right) =\left\{ 
\begin{array}{c}
0,\ t<0 \\ 
1,\ t>1.%
\end{array}%
\right. 
\]

\subsection{Solving the Problem for a general initial state}
\noindent

Here, we present the control Hamiltonian which solves the \emph{Problem}
for an arbitrary initial state with purity greater than a non-zero coherence, \textit{viz.,}%
\begin{equation}
\rho \left( 0\right) =\frac{1}{2}I+\frac{1}{2}\left( v_{x}\left( 0\right)
\sigma _{x}+v_{y}\left( 0\right) \sigma _{y}+v_{z}\left( 0\right) \sigma
_{z}\right) ,
\end{equation}
where
\[
\ 0 < c=v_{x}\left( 0\right) ^{2}+v_{y}\left( 0\right) ^{2}<\
p=v_{x}\left( 0\right) ^{2}+v_{y}\left( 0\right) ^{2}+v_{z}\left( 0\right)
^{2}.
\]
Now, we define the unitary operator%
\[
U_{\theta }:=\left( 
\begin{array}{cc}
e^{-i\theta /2} & 0 \\ 
0 & e^{i\theta /2}%
\end{array}%
\right) =\left( \cos \frac{\theta }{2}\right) I-i\left( \sin \frac{\theta }{2%
}\right) \sigma _{z},
\]%
where $\theta \in \left[ 0,2\pi \right) $ is such that%
\[
v_{x}\left( 0\right) =\sqrt{c}\cos \theta \ ,\ v_{y}\left( 0\right) =\sqrt{c}%
\sin \theta .
\]%
Using $U_{\theta }$, we define the following state which has the previous special form as well as the same purity
and coherence of $\rho \left( 0\right) $:%
\[
\tilde{\rho}\left( 0\right) :=U_{\theta }^{\ast }\rho \left( 0\right)
U_{\theta }=\frac{1}{2}I+\frac{1}{2}\sqrt{c}\sigma _{x}+\frac{1}{2}%
v_{z}\left( 0\right) \sigma _{z}.
\]%
Now, let $\tilde{H}\left( t\right)$ be the control Hamiltonian that solves the \textit{Problem} for $\tilde{\rho}\left( 0\right)$ and time span $T>0$. Since $U_{\theta }$ is constant and commutes with $\sigma _{z}$, the control Hamiltonian which solves the \textit{Problem} for $\rho\left( 0\right)$ and time span $T>0$ is given by:\footnote{See the explicit expression in Eq. (\ref{ucontrol_econtrol-hamiltonian}) below.}
\[
H\left( t\right) :=U_{\theta } \tilde{H}\left( t\right) U_{\theta }^{\ast }.
\]

\begin{remark}
Naturally, a control Hamiltonian which solves the \emph{Problem} for a general initial state must be unitarily-equivalent to the control Hamiltonian which solves the \emph{Problem} for some special initial state, because general states are related to the special ones by a change of variables (specifically, a suitable rotation in the $xy$-plane).
\end{remark}

\subsection{Algorithm}
\noindent
\begin{quotation}
\label{ucontrol-pd_algorithm} 
To solve the \emph{Problem} for the initial state
\[
\rho \left( 0\right) =\frac{1}{2}I+\frac{1}{2} \left( v_{x}\left( 0\right) \sigma
_{x}+v_{y}\left( 0\right) \sigma_{y}+v_{z}\left( 0\right) \sigma _{z} \right)
\]
with
\[
0< c = v_{x}^{2}\left( 0\right)+v_{y}^{2}\left( 0\right) < p = c + v_{z}^{2}\left( 0\right),
\]
do:

i)\ Set $\epsilon := - signal\left( v_{z}\left( 0\right) \right) $ and $\theta \in \left[ 0,2\pi \right) $ such that
\[
v_{x}\left( 0\right) =\sqrt{c}\cos \theta \ ,\ v_{y}\left( 0\right) =\sqrt{c}%
\sin \theta ;
\]

ii) Solve the following system for $u$, $\Delta t_{1}$ and $\Delta t_{3}$:%
\begin{equation}
\left\{ 
\begin{array}{l}
\tan \left( \frac{1}{4}\sqrt{4u^{2}-\gamma ^{2}}\Delta t_{1}\right) =\frac{%
\sqrt{4u^{2}-\gamma ^{2}} \sqrt{c} }{\gamma \sqrt{c}
+2 u  \sqrt{p-c} } \\ 
\  \\ 
\sin \left( \frac{1}{4}\sqrt{4u^{2}-\gamma ^{2}}\Delta t_{3}\right) =\frac{%
e^{\gamma \left( \Delta t_{1} + \Delta t_{3}\right)/4}\sqrt{4u^{2}-\gamma ^{2}} \sqrt{c} }{2%
\sqrt{u^{2}p + \gamma u \sqrt{c} \sqrt{p-c} }} \\ 
\  \\ 
u>\gamma /2,\ \Delta t_{1}>0,\ \Delta t_{3}>0,\ \Delta t_{1}+\Delta
t_{3}\leq T;%
\end{array}%
\right.   \label{ucontrol_solution-system}
\end{equation}%
(Alternatively, one can prescribe a positive value for $u$, determine $\Delta
t_{1}$ and $\Delta t_{3}$ from the first and second equations of System (%
\ref{ucontrol_solution-system}) and then verify if the inequalities are also
satisfied.)

\medskip 

iii)\ Define%
\[
\Delta t_{2}:=T-\Delta t_{1}-\Delta t_{3}\geq 0; 
\]

iv) Define the control Hamiltonian by:%
\[
H\left( t\right) 
=-\epsilon u\left[ \mathfrak{h}(\Delta t_{1}-t)-\mathfrak{h}%
(t-\Delta t_{2})\right] \left[ \left( \sin \theta \right) \sigma _{x}-\left(
\cos \theta \right) \sigma _{y}\right], \ 0 \leq t \leq T .
\label{ucontrol_econtrol-hamiltonian}
\]
\end{quotation}

\begin{remark}
The System of equations (\ref{ucontrol_solution-system}) has solutions for $u>\gamma /2$
sufficiently large (implying that $\Delta t_{1}$ and $\Delta t_{3}$ are
correspondingly small). To verify, we note the following approximations
valid under such conditions:%
\[
\Delta t_{1}\approx \frac{2}{u}\arctan \left( \left\vert \frac{v_{x}\left(
0\right) }{v_{z}\left( 0\right) }\right\vert \right) , 
\]%
\[
\left\vert v_{z}\left( \Delta t_{1}\right) \right\vert \approx e^{-\gamma
\Delta t_{1}/4}\sqrt{v_{z}^{2}\left( 0\right) +v_{x}^{2}\left( 0\right) }%
, 
\]%
\[
\left\vert v_{x}\left( 0\right) \right\vert \approx e^{-\gamma \Delta
t_{3}/4}\left\vert v_{z}\left( \Delta t_{1}\right) \right\vert \sin \left( 
\frac{u}{2}\Delta t_{3}\right) . 
\]%
The first equation gives an approximation for $\Delta t_{1}$, the second
equation implies $\left\vert v_{z}\left( \Delta t_{1}\right) \right\vert
>\left\vert v_{x}\left( 0\right) \right\vert $ and the third equation has a
sine function which oscillates very quickly; therefore, for relatively small
values of $\Delta t_{3}$ it follows that $e^{-\gamma \Delta t_{3}/4}\approx
1 $ and\ 
\[
\Delta t_{3}\approx \frac{2}{u}\arcsin \left( \frac{\left\vert v_{x}\left(
0\right) \right\vert }{\sqrt{v_{z}^{2}\left( 0\right) +v_{x}^{2}\left(
0\right) }}\right) . 
\]
\end{remark}

\subsection{A numerical example}
\noindent
Let us illustrate the application of the \emph{Solution} using numerical
values presented in \cite{LS2005}.

Consider a system with damping coefficient $\gamma =0.1$ and assume the initial state has purity $p=0.8$ and coherence $c=0.3$,
\[
\rho \left( 0\right) =\frac{I}{2}+\frac{\sqrt{0.3}}{2}\sigma _{x}+\frac{%
\sqrt{0.5}}{2}\sigma _{z}. 
\]

In this case, the breakdown time is%
\[
t_{b}=\frac{p-c}{\gamma c}\approx 16.67. 
\]

If we set $u=0.2$, then the system of equations (\ref{ucontrol_solution-system}) implies%
\[
\Delta t_{1}\approx 5.79,\ \ \Delta t_{3}\approx 9.11.
\]%
Then, $\Delta t_{1}+\Delta t_{3}=14.90$. For $T=20>t_{b}$, the application
of our control strategy gives the following results: \emph{the purity
evolves from the initial value }$0.8$\emph{\ to the final value }$\approx
0.63$\emph{; the coherence evolves from the initial value }$0.3$\emph{\ to
the final (and same) value }$0.3$\emph{, decreasing to zero during the first
stage (between }$t=0$\emph{\ and }$t\approx 5.8$)\emph{, staying equals to
zero\ during the second stage (between }$t\approx 5.8$\emph{\ and }$%
t\approx 10.9$)\emph{\ and increasing to }$0.3$\emph{\ during the third
stage (between }$t\approx 10.9$\emph{\ and }$t=20$).

Figure 1 gives the graph of the $y$-component of the
control Hamiltonian, the path of the Bloch vector in the $xz$-plane during
the control process and the graph of purity and coherence as functions of
time:

\begin{figure} [htbp]
\centerline{\epsfig{file=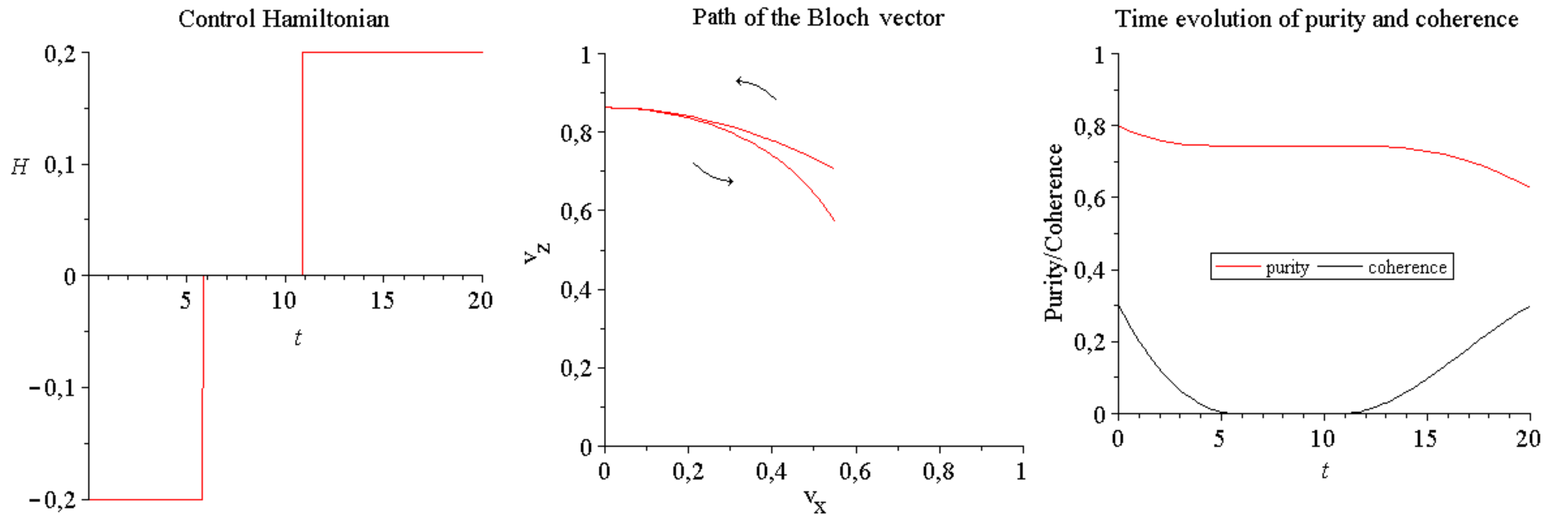, width=12.cm}} 
\vspace*{13pt}
\fcaption{\label{ucontrol_fig1}Example of controlling the coherence in the dephasing qubit.}
\end{figure}

\section{Conclusions\label{section_conclusion}}
\noindent

Characteristic of many problems in control theory is the need to develop idiosyncratic strategies -- even for situations in which there are general procedures to solve them, because the advantages of a specific procedure may be worthwhile in a particular application. We think this fact is well illustrated here by our open-loop strategy to control the coherence in the
dephasing qubit. For a comparison with the tracking-control strategy of 
\cite{LS2005}, we remark that: the tracking-control can be applied to stabilize the coherence only for a time span smaller than the breakdown time (with energetic expenditure reducing as the control period decreases), while our strategy can be applied for any prescribed time span (with the control fields becoming larger as the control period decreases). The trick of our control strategy lies in the first and third stages, which must be performed as quickly as necessary since purity decreases during them; for this strategy to be successful, the control field's intensity must reach sufficiently large values, as one can deduce from Eq. (\ref{ucontrol_dynamics-3_t1}).

We believe the reasoning presented here can be naturally adapted to control the coherence of other Markovian quantum systems having a \emph{decoherence-free subset of states}, with the help of a \emph{coherent vector representation}. The \textit{Strategy} (Sec. \ref{section_strategy}) is general, in the sense that it may be applied to recover the coherence in models other than the \emph{dephasing qubit}; nevertheless, the first and third stages must be carried out taking into account specific details of each model. A natural development of this work is the application of the \textit{Strategy} in more complex and realistic situations, what can be more interesting and more useful, but more laborious too.

Turning to the important question about the energy expenditure of the control
process, we close the paper stating a new problem:

\begin{quotation}
\textbf{Optimal Control Problem}: In the \emph{dephasing qubit}, for a given $T>0$ and an initial state $\rho \left( 0\right) $, set a control Hamiltonian to steer the
state's evolution according with the dynamics so that (i) the coherence of the system' state after the time span $T$ be equal to the coherence
of the initial state and (ii) the expenditure of energy in the process is minimum, with this expenditure being defined by a quadratic form on the control fields \cite{DAD2001, RD2007}, \textit{e.g.},
\[
K_{u}=\int_{0}^{T}\left( u_{1}^{2}\left( t\right) +u_{2}^{2}\left( t\right)
+u_{3}^{2}\left( t\right) \right) dt.
\]
\end{quotation}


\appendix{ \textit{The limit time}}
\noindent
For the \textit{dephasing qubit}, we define the \textquotedblleft \emph{limit time%
}\textquotedblright\ by the maximum time span $\tilde{T}$ that one can spend
on steering an initial state to the decoherence-free subset of states $V_{z}$ and,
after, to some final state having coherence equals to the initial value,
using solely a control Hamiltonian as given by Eq. (\ref{model-pd_control-hamiltonian}). Specializing this definition for Hamiltonians having the shape (\ref{ucontrol_econtrol-hamiltonian}), we define the \textquotedblleft \emph{limit
control field}\textquotedblright\ as the minimal value that a
(constant) control field $u$ can assume in the solutions of the system (\ref%
{ucontrol_solution-system}) when $T=\tilde{T}$.

The relevance of these concepts is the following: for a control period $T\geq \tilde{T}$, the control of coherence can be done using a control field $u = \tilde{u}$, while for a control period $T < \tilde{T}$, the control field must satisfy $u > \tilde{u}$.

Quantities $\tilde{T}$ and $\tilde{u}$ are mutually dependent and are characterized by the property that \emph{the purity of the initial state is
fully reduced to the initial value of the coherence at the end of the
corresponding control process}, namely: 
\begin{equation}
v_{z}\left( \tilde{T}\right) =0.  \label{ucontrol_mcondition}
\end{equation}

To calculate $\tilde{T}$  and $\tilde{u}$, we combine the two equations of the system (\ref{ucontrol_solution-system})
with condition (\ref{ucontrol_mcondition}); after some algebraic
manipulation, we get the following system for $\tilde{u}$ and $\Delta \tilde{%
t}_{1}$ and $\Delta \tilde{t}_{3}$, where $\tilde{T}=\Delta \tilde{t}%
_{1}+\Delta \tilde{t}_{3}$:\footnote{%
We remark that Eq. (\ref{ucontrol_mcondition}) is equivalent to the
third equation of System (\ref{ucontrol_mconditionre}).}
\begin{equation}
\left\{ 
\begin{array}{l}
\Delta \tilde{t}_{1}=\frac{4}{\sqrt{4\tilde{u}^{2}-\gamma ^{2}}}\arctan
\left( \frac{\sqrt{4\tilde{u}^{2}-\gamma ^{2}}v_{x}\left( 0\right) }{\gamma
v_{x}\left( 0\right) -2\epsilon \tilde{u}v_{z}\left( 0\right) }\right)  \\ 
\  \\ 
\Delta \tilde{t}_{3}=\frac{4}{\sqrt{4\tilde{u}^{2}-\gamma ^{2}}}\arcsin
\left( \frac{e^{\gamma \left( \Delta \tilde{t}_{1}+\Delta \tilde{t}%
_{3}\right) /4}\sqrt{4\tilde{u}^{2}-\gamma ^{2}}\left\vert v_{x}\left(
0\right) \right\vert }{2\sqrt{p_{0}\tilde{u}^{2}-\epsilon \gamma v_{z}\left(
0\right) v_{x}\left( 0\right) \tilde{u}}}\right)  \\ 
\  \\ 
\tan \left( \frac{\sqrt{4\tilde{u}^{2}-\gamma ^{2}}}{4}\Delta \tilde{t}%
_{3}\right) =-\frac{\sqrt{4\tilde{u}^{2}-\gamma ^{2}}}{\gamma }.%
\end{array}%
\right.   \label{ucontrol_mconditionre}
\end{equation}

Since this system is very complicated, it is useful to know that $\tilde{T}$ is  greater then the breakdown time, given by Eq. (\ref{model_pd_breakdown-time}). This fact is easy to prove and it implies a super estimation of $\tilde{u}$, to which we now turn (with some omissions in the argument). Using
that $\Delta t_{1}$ and $\Delta t_{3}$ are decreasing functions of $u$, a
sub estimation of $\Delta \tilde{t}_{3}$ implies a super estimation of $\tilde{%
u}$; since, in general, $\Delta t_{3}\geq \Delta t_{1}$ and $\Delta
t_{1}+\Delta t_{3}>t_{b}$, it follows 
\[
\Delta \tilde{t}_{3}>\frac{t_{b}}{2}=\frac{p-c}{2\gamma c}
\]%
Inserting this sub estimation\ for $\Delta \tilde{t}_{3}$ in the third
equation of the System (\ref{ucontrol_mconditionre}), we conclude that \emph{%
the minimal control field }$\tilde{u}$\emph{\ is not greater than the
solution }$\xi $\emph{\ of the following equation}:%
\begin{equation}
\tan \left( \sqrt{4\xi ^{2}-\gamma ^{2}}\frac{p-c}{8\gamma c}\right) =-\frac{%
\sqrt{4\xi ^{2}-\gamma ^{2}}}{\gamma }.  \label{ucontrol_mminest}
\end{equation}

\end{document}